\documentclass[12pt,preprint]{aastex}

\slugcomment{To be published in ApJ}

\shorttitle{Pulse Profile Change and Glitch in 4U 0142+61}
\shortauthors{Morii and Kawai}

\def\Gtsim{\mathrel{\vbox to 0pt{\hbox{$\sim$}}\llap{$>$}}}

\begin{document}


\title{Pulse Profile Change Possibly Associated with a Glitch
	 in an Anomalous X-ray Pulsar 4U 0142+61}


\author{Mikio Morii\altaffilmark{1}}
\affil{Japan Aerospace Exploration Agency,
        Sengen 2-1-1, Tsukuba-shi, Ibaraki-ken 305-8505, Japan}
\altaffiltext{1}{Department of Physics, Tokyo Institute of Technology,
	Ookayama 2-12-1, Meguro-ku, Tokyo 152-8551, Japan}
\email{morii@oasis.tksc.jaxa.jp}

\author{Nobuyuki Kawai}
\affil{Department of Physics, Tokyo Institute of Technology,
	Ookayama 2-12-1, Meguro-ku, Tokyo 152-8551, Japan}

\and

\author{Noriaki Shibazaki}
\affil{Department of Physics, Rikkyo University,
  Nishi-Ikebukuro, Tokyo 171-8501, Japan}




\begin{abstract}
We report a glitch-like pulse frequency deviation
from the simple spin-down law in an anomalous X-ray pulsar (AXP)
4U 0142+61 detected by {\it ASCA} observations.
We also found a significant pulse profile change
after the putative glitch.
The glitch parameters resemble those found in another
AXP 1RXS J170849.0$-$400910,
in the Vela pulsar, and in other radio pulsars.
This suggests that the radio pulsars and AXPs have the
same internal structure and glitch mechanism.
It must be noted, however, that the pulse frequency anomaly can also
be explained by a gradual change of the spin-down rate ($\dot{P}$)
without invoking a glitch.
\end{abstract}




\section{Introduction}

Anomalous X-ray pulsars (AXPs) are a small group of X-ray emitting pulsars
(see \cite{Mereghetti 2002} for a review).
They are not likely to be accretion-powered pulsars
with a binary companion,
since the fluxes of the optical counterparts are
too small for high-mass binary companions or the Doppler modulation
due to the binary motion has not yet been found.
They are also not rotation-powered pulsars,
since the observed X-ray luminosities
($L_{\rm X} \sim 10^{34} - 10^{36}$ ergs s$^{-1}$)
exceed the spin-down energy loss rates of neutron stars
($\dot{E} = 4\pi^2 I\dot{P}/{P}^3 \sim 10^{32.6}$ ergs s$^{-1}$).

Accretion-powered pulsars without a binary companion
(fossil accretion disk) \citep{Chatterjee 2000, Alpar 2001}
and those with a small binary companion
\citep{Mereghetti-Stella 1995}
have been discussed.
However, the pulse timing that originates in the accretion disk
(large timing noises and/or persistent spin-up periods)
is not found.
AXP 1E 2259+586 \citep{Kaspi 2003}
and 1RXS J170849.0$-$4000910
\citep{Kaspi-Lackey-Chakrabarty 2000, Kaspi-Gavriil 2003}
exhibited glitches
similar to those of rotation-powered pulsars.
The timing behavior of AXPs is similar to
a solitary neutron star with no accretion disk.
AXPs are also characterized by
soft X-ray spectra ($\Gamma \Gtsim 2$), no radio detections,
slow rotation periods ($5 - 12$ s), and burst activities.

\citet{Thompson-Duncan 1996, Thompson-Duncan 2001}
proposed the novel hypothesis
that SGRs (soft gamma repeaters) and AXPs are solitary neutron stars
with ultra-strong magnetic fields ($10^{14} - 10^{15}$ G),
referred to as ``magnetars.''
X-ray photons are produced in this model by the release of
the strong magnetic field energy stored in the
neutron star crust, or a twisted internal magnetic field of neutron stars.
Recent discoveries of burst activities
\citep{Gavriil-Kaspi-Woods 2002, Kaspi 2003},
an optical pulsation \citep{Kern-Martin 2002}, and
cyclotron spectral features 
\citep{Gavriil-Kaspi-Woods 2002, Ibrahim 2002, Ibrahim-Swank-Parke 2003,
Rea 2003}
support the magnetar model.
Interestingly, 
the glitch and the pulse profile change were observed simultaneously
during the outburst of 1E 2259+586
\citep{Kaspi 2003}.

The X-ray source 4U 0142+61 was discovered by {\it Uhuru};
its pulsation period of $8.7$ s was discovered by
{\it EXOSAT} \citep{Israel-Mereghetti-Stella 1994}.
4U 0142+61 was one of the prototypes of anomalous X-ray pulsars 
\citep{Hellier 1994, Mereghetti-Stella 1995, vanParadijs 1995}.
A discovery of the optical pulsation
supports the magnetar scenario by suggesting
that the optical emission comes from the magnetosphere
of the neutron star, not from the accretion disk
\citep{Kern-Martin 2002}.

The long-term stability of its intensity, spectrum, and pulse profile
were reported through a comparison between
{\it ASCA} observations in 1994 and in 1998 \citep{Paul 2000}.
The high stability over 4.4 yrs was reported
following the long-term monitoring observations of {\it RXTE}
from Nov. 1996 to Apr. 2001 \citep{Gavriil-Kaspi 2002},
and the pulse frequency and its derivative were precisely determined.
However, this {\it RXTE} observation was not complete
due to a 2.0-yr gap from May 1998 to Mar. 2000,
dividing its total 4.4-yr span into two spans,
the ``First'' (1.3 yr) and  the ``Second'' (1.1 yr)
(Table 2 in \cite{Gavriil-Kaspi 2002} or
Fig. \ref{fig: Spin-down of 4U 0142+61}).
\cite{Gavriil-Kaspi 2002} combined these spans and
provided the unified ephemerides,
``A'' and ``B,'' which are
qualitatively similar (Table 2 in \cite{Gavriil-Kaspi 2002}).

The {\it ASCA} observations in 1998 and 1999
were performed just during the gap between
the {\it RXTE} observations.
Those in 1999 in particular were carried out over a period of a month,
and therefore precise determination of the pulse frequency 
was possible.
We discovered in our study that the pulse frequency of 4U 0142+61
in the 1999 {\it ASCA} observations was significantly higher than 
those predicted by the unified ephemerides (A, B) and the first ephemeris,
and it is also marginally higher than the prediction by the second ephemeris.
These deviations exclude the unified ephemerides and
suggest the presence of a glitch within
the 2.0-yr gap before the 1999 {\it ASCA} observations.
In addition, we detected a morphological change of the X-ray pulse, in which
the pulse profile after the glitch differed significantly from
those before the glitch.

\section{Observations}

4U 0142+61 was observed with {\it ASCA} eight times during
its mission life.
The observations were undertaken in 1994, 1998, and 1999.
This source was monitored over a period of
about one month with six exposures during the last term.
These observations are summarized in Table \ref{table: ASCA obs}.
All of the observations were taken by GIS and SIS detectors.
Almost all GIS observations were taken
in the PH mode with 1024 PHA bins and timing bit = 0
at high or medium bit rate,
in which the time resolutions were 62.5 or 500 ms.
We analyzed only the GIS-2 and GIS-3 data
with high and medium bit rates.

\begin{table}
\begin{center}
\caption{Summary of ASCA Observations}
\begin{tabular}{lcclc}
\hline\hline
\multicolumn{1}{c}{Start Date}& \multicolumn{1}{c}{Exposure} &
\multicolumn{1}{c}{Reference Epoch} & \multicolumn{1}{c}{$\nu$} & \multicolumn{1}{c}{$\dot{\nu}$}  \\
\multicolumn{1}{c}{(UT)}  & \multicolumn{1}{c}{(ks)} &
\multicolumn{1}{c}{(MJD)} & \multicolumn{1}{c}{(Hz)} & \multicolumn{1}{c}{(Hz s$^{-1}$)} \\
\hline
1994 Sep 18 (21:35:39) & 18.4 & 49614.17288 & 0.1151016(8)   & $\cdots$ \\
1998 Aug 21 (11:23:48) & 18.9 & 51046.69875 & 0.1150972(6)   & $\cdots$ \\
\hline
1999 Jul 29 (16:22:15) & 23.2 &             &  &\\
1999 Aug 03 (20:03:29) & 23.5 &             &  &\\
1999 Aug 09 (18:37:56) & 11.0 &             &  &\\
1999 Aug 12 (22:24:03) & 19.4 & 51403.13258 & 0.115097645(4) &
							 $-2.4^{+0.8}_{-0.7} \times 10^{-14}$ \\
1999 Aug 21 (19:06:08) & 25.1 &             &  &\\
1999 Aug 27 (03:01:19) & 18.1 &             &  &\\
\hline
\multicolumn{5}{l}{Reference epochs are given at the mid-observation.
 Numbers in parentheses are 1 $\sigma$ uncertainties.} \\
\end{tabular}
\label{table: ASCA obs}
\end{center}
\end{table}

We obtained all the {\it ASCA} observation data for 4U 0142+61 from 
the High Energy Astrophysics Science Archive Research Center (HEASARC)
archive site.\footnote{ftp://legacy.gsfc.nasa.gov/FTP/asca/data/rev2/}
We used the SCREENED data, which had been subjected
to the standard screening procedure (Revision 2).
We selected the source photons from a circular region
with a radius of 5.9 arcmin
centered on the position with the pulsar peak counts,
and we used events with an energy
of 0.5 to 10 keV (44$-$848 in PI space).
The source flux was constant during all observations.
The event arrival times were corrected to the value at 
the solar system barycenter using \texttt{timeconv v1.53} (HEASOFT v5.2).
We incremented the times by half of their time resolutions
to mix the high and medium bit rate data and compensate for the difference 
of the time resolutions,
since the times recorded were
the leading edge of the time duration
of the resolution (M.~Hirayama; private communication).

\section{Timing Analysis}

We proceeded with the following steps
to determine the pulse frequencies ($\nu$) and their derivatives ($\dot{\nu}$)
for the observation terms of 1994, 1998, and 1999,
on the assumption that the higher order derivatives were zero.
We calculated both values at the reference epochs ($t_0$),
which were selected at the middle of the terms,
to minimize the effect of the correlated error
between $\nu$ and $\dot{\nu}$.
Hereafter, a subscript ``$0$'' denotes a parameter evaluated
at the reference epoch $t_0$.

\subsection{Epoch-folding Method}
The rough ($\nu_0$, $\dot{\nu_0}$) pairs were estimated
by the epoch-folding method,
in which the best ($\nu_0$, $\dot{\nu_0}$) pair was found to maximize
the quantity $S$ from each folded light curve:
\begin{eqnarray}
S & = & \sum^n_{j = 1} \frac{[N_j - <N_j>]^2}{\sigma_{(N_j - <N_j>)}^2} \\
  & = & \sum^n_{j = 1} \frac{[N_j - <N_j>]^2}{(1 - 2/n) N_j + N/n^2},
\end{eqnarray}
where $n = 16$ is the number of bins in one period, $N_j$ ($j = 1,2, \cdots, n$) is
the number of events in the $j$-th bin,
$N = \sum^n_{j = 1} N_j $ is the total number of events in the observation term,
and $<N_j> = N/n$ is the mean number of events for the bins.

\subsection{Pulse Arrival Time Analysis \label{section: pulse number counting}}
More accurate determinations of the pulse frequency 
and its uncertainty were achieved by 
counting the pulse numbers and the pulse arrival times.
The pulse numbers $\phi_{\rm tot}$ at any time $t$ can be expressed by
the Taylor expansion around the reference epoch $t_0$,
as in \citet{Gavriil-Kaspi 2002}.
We divided each of the observation terms into short segments.
The time series was split
into short segments of about 30 min duration every $\sim 96$ min
for the terms of 1994 and 1998.
Each of the six observations for the term of 1999
(Table \ref{table: ASCA obs})
corresponds to a segment.
The pulse profiles of the segments were made by folding the time series with
the best ($\nu_i$, $\dot{\nu_i}$) values of the $i$-th segment 
($i = 1,2, \cdots, n_{\rm seg}; n_{\rm seg}$: the number of segments)
obtained by the epoch-folding method cited above,
in which the reference ``phase zero'' epoch of each pulse profile was
taken at the mid-time of each segment.
The template pulse profile of the term
was generated in the same way using all the segments.
Both the template and the segment pulse profiles were
cross-correlated in the Fourier domains;
the Fourier components equal to or above the fifth order were discarded.
Thus, we obtained the phase offset ($\Delta N_i$)
of each segment from the template. The uncertainties of the
offsets were estimated
by Monte Carlo simulation,
adding the Poisson errors to the pulse phase bins.

We were then able to calculate ($\nu_0$, $\dot{\nu_0}$) most accurately
by imposing a constraint on the segment pulses
in which the cumulative pulse phase $\phi_{\rm tot}(t)$
of the $i$-th segment relative to the template
must be an integer ($N_i$) plus the phase offsets ($\Delta N_i$).

The best combination of the $N_i (i = 1,2, \cdots, n_{\rm seg})$
and ($\nu_0$, $\dot{\nu}_0$) pair
was found to minimize the following $\chi^2$ value:

\begin{equation}
\chi^2(\nu_0, \dot{\nu}_0; N_1, N_2, \cdots, N_{n_{\rm seg}}) 
 = \sum_i \frac{[\phi_{\rm tot}(t_i; \nu_0, \dot{\nu}_0) - (N_i + \Delta N_i)]^2}{\sigma_{\Delta N_i}^2}
\end{equation}
Here, $N_i$ was selected within the error of $\phi_{\rm tot}(t_i)$
induced from the uncertainties of ($\nu_0$, $\dot{\nu}_0$)
roughly determined by the epoch-folding method.
The uncertainties of ($\nu_0$, $\dot{\nu}_0$) can be
calculated appropriately,
because this analysis uses the $\chi^2$ fitting method.

\section{Results}
\subsection{Pulse Frequency History}
\begin{figure}
 \begin{center}
 \includegraphics[height=8cm]{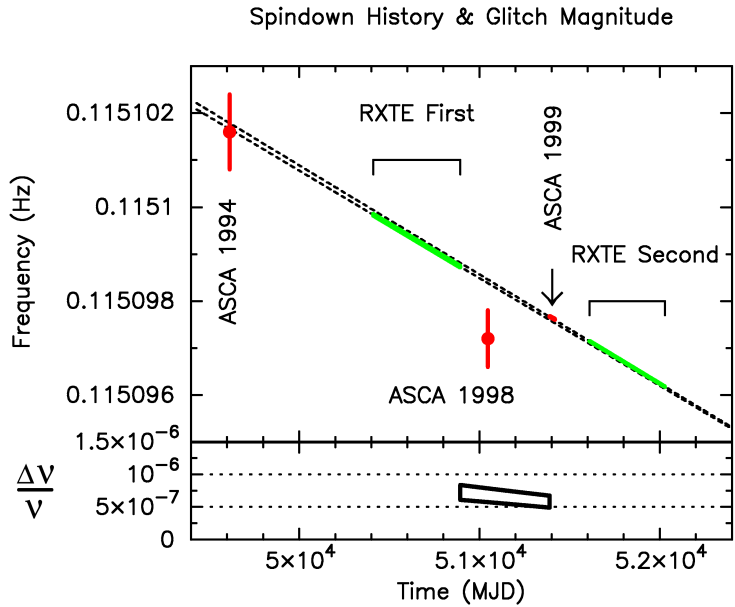}
\caption{(Upper panel:) The spin-down history of 4U 0142+61.
The horizontal and vertical axes are the time (MJD) and pulse frequency (Hz).
The {\it RXTE} ``First'' and ``Second'' ephemerides are indicated
with solid lines.
The extrapolations of these ephemerides are designated by dotted lines.
The {\it ASCA} ephemeris in 1999 are represented by a short solid line
indicated by an arrow.
The frequency obtained by the 1994 and 1998 {\it ASCA} observations are
indicated in bars with $1 \sigma$ errors.
(Lower panel:) The glitch magnitude ($\frac{\Delta \nu}{\nu}$)
when the glitch occurred
at the time (MJD) is designated as a $1 \sigma$ error region.}
\label{fig: Spin-down of 4U 0142+61}
 \end{center}
\end{figure}

\begin{figure}
 \begin{center}
 \includegraphics[height=8cm]{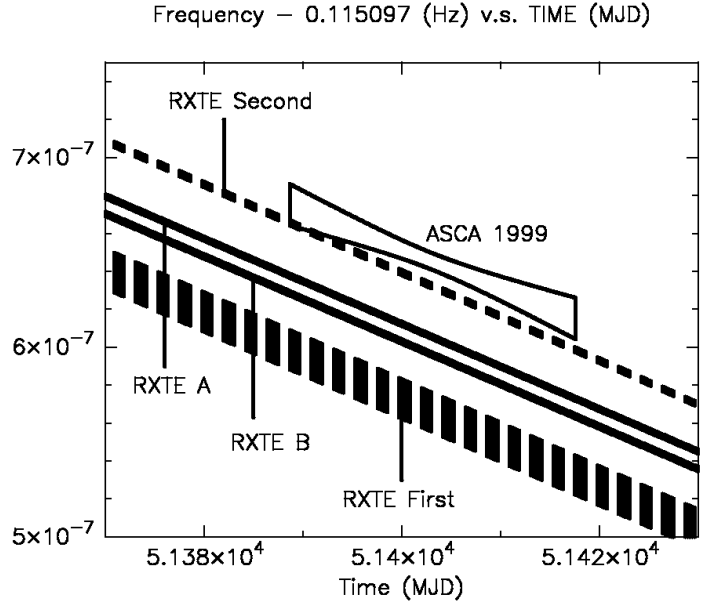}
\caption{
Frequency history of 4U 0142+61 around the 1999 {\it ASCA} observation
(close-up of Fig. \ref{fig: Spin-down of 4U 0142+61}).
The horizontal and vertical axes are the time (MJD)
and the frequency minus 0.115097 (Hz).
The spin-down history of 4U 0142+61 expected by the {\it RXTE}
observations are indicated by the thick lines, where the widths of
the lines correspond to $1 \sigma$ uncertainties.
The unified ephemerides (``A'' and ``B'') are represented by solid lines:
the extrapolations from the separate ephemerides 
(``First'' and ``Second'') are indicated by the striped lines.
The error region in the {\it ASCA} observation in 1999 is also depicted.}
\label{fig: timing_check_large2}
 \end{center}
\end{figure}

The pulse frequencies and their derivatives obtained by {\it ASCA} observations
are listed in Table \ref{table: ASCA obs}.
The observations in 1994 and 1998 were
too short to determine the $\dot{\nu}$,
and therefore we calculated only the $\nu$ by setting $\dot{\nu} = 0$ for these terms.
The spin-down history is provided in Fig. \ref{fig: Spin-down of 4U 0142+61}.
The frequencies in the {\it ASCA} observations in 1994 and in 1998 were consistent with
all of the {\it RXTE} ephemerides \citep{Gavriil-Kaspi 2002}.
However, the frequency for that in 1999 was significantly higher than 
the unified ephemerides (A or B) and the First ephemeris, and marginally higher
than the Second ephemeris of {\it RXTE} (Fig. \ref{fig: timing_check_large2}).
The frequencies at the reference epoch of the 1999 {\it ASCA} observation
expected by the {\it RXTE} ephemerides \citep{Gavriil-Kaspi 2002}
are listed in Table \ref{table: ASCA nu deviation}.
The same table also reveals the probability of the null hypothesis
that the frequency of the {\it ASCA} observation and those of
the {\it RXTE} are the same.

\begin{table}
\begin{center}
\begin{tabular}{clrr}
\hline\hline
Ephemeris & $\nu$ (Hz) & $\sigma$ & Probability \\
\hline

First span only  & 0.115097566(10)  &  7.0 & $ 9.7 \times 10^{-13}$ \\
Second span only & 0.1150976321(14) &  2.8 & $ 2.4 \times 10^{-3} $ \\
Combined ``A''   & 0.1150976052(12) &  8.8 & $ < 10^{-13} $ \\
Combined ``B''   & 0.1150975960(13) & 10.7 & $ < 10^{-13} $ \\
\hline
{\it ASCA}   & 0.115097645(4)   & & \\
\hline
\hline
\multicolumn{4}{l}{Numbers in parentheses are $1 \sigma$ uncertainties.}
\end{tabular}
\caption{Expected frequency at the reference epoch of
the 1999 {\it ASCA} observation (51403.13258 MJD)
from the {\it RXTE} ephemerides.
The significance of the deviation from the {\it ASCA} ephemeris
is also indicated.
\label{table: ASCA nu deviation}}
\end{center}
\end{table}

\subsection{Pulse Profile}
We searched for the pulse profile changes.
We first compared the pulse profiles of every segment 
among the observations in 1999 with
their template pulse profiles.
The pulse phases of the segment pulse profiles were aligned with
the template using the same cross-correlation procedure used in the
pulse arrival time analysis ({\S} \ref{section: pulse number counting}).
Each segment pulse profile ($D$) was fitted to
the template pulse profile ($T$)
to minimize the reduced $\chi^2$ statistic by adjusting parameters $a$ and $b$
as follows.
\begin{equation}
\chi^2_\nu = \frac{1}{n - 2}
	 \sum^{n}_{j = 1}
	 \frac{[D_j - (a \times T_j - b)]^2}
	{\sigma_{D_j}^2}.
\label{eq: chi2_r}
\end{equation}
Here, $n = 16$ is the number of phase bins, and
$D_j$($T_j$) is the number of events in the $j$-th bin
of the segment (the template pulse profiles).
The number of degrees of freedom is $\nu = n - 2$.

\begin{figure}
 \begin{center}
 \includegraphics[height=12cm, angle=270]{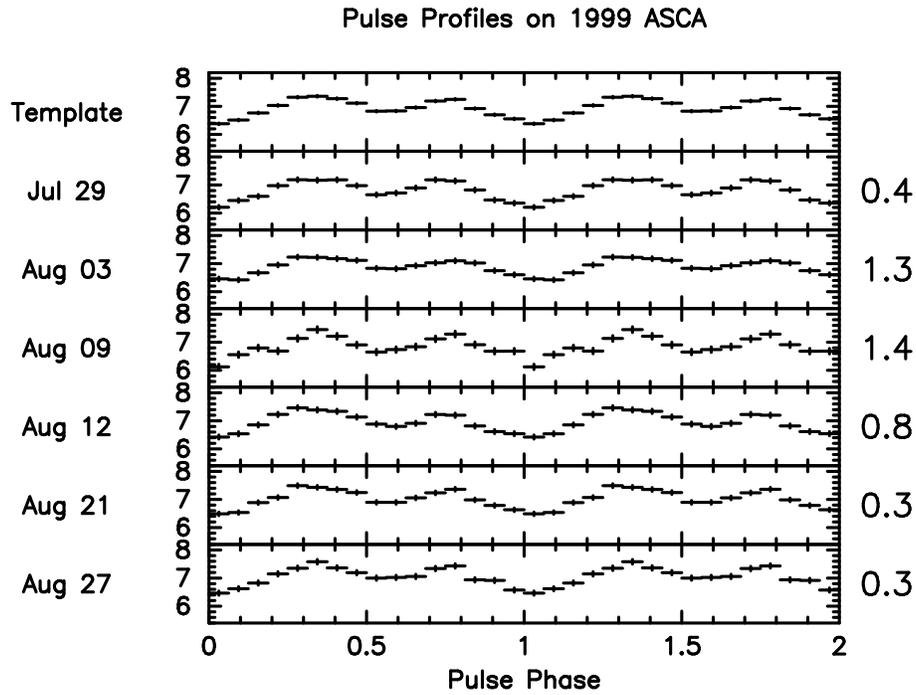}
\caption{Pulse profiles of 4U 0142+61 in the 1999 ASCA observations.
The vertical and horizontal axes are shown
in units of the count rates (counts s$^{-1}$) and pulse phases up to 2 periods.
 (The top panel:) The template pulse profile in the term of
the 1999 {\it ASCA} observation.
(The second to the bottom panel:) The segment pulse profiles
of the 1999 {\it ASCA} observations.
The reduced $\chi^2$ values (see text) are indicated
in the right of the panels.
There are no significant pulse profile changes
during the 1999 {\it ASCA} observation.}
\label{fig: 1999 ASCA pulse profile}
 \end{center}
\end{figure}

\begin{figure}
 \begin{center}
 \includegraphics[height=10cm, angle=270]{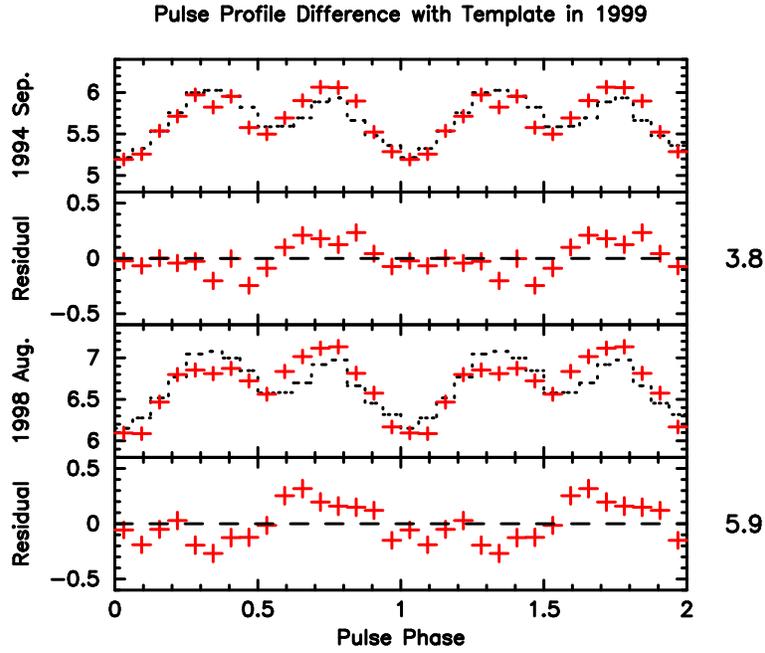}
\caption{The template pulse profiles in Sep. 1994 and Aug. 1998
are shown in the upper and lower panels.
The vertical and horizontal axes are shown in units of the count rates
(counts s$^{-1}$) and pulse phases up to 2 periods.
The template pulse profile in 1999 is also shown by dotted pulse profiles,
where the normalizations and baseline offsets were adjusted
to minimize the reduced $\chi^2$ (Equation \ref{eq: chi2_r}).
 The pulse profiles in
1994 and 1998 differed significantly from that in 1999
with $\chi^2_\nu$ of $3.8$ and $5.9$.
The residuals from adjusted 1999 template pulse profiles are also shown.
}
\label{fig: 1994 1998 ASCA pulse profile}
 \end{center}
\end{figure}

There were no significant variations
 ($\chi^2_\nu \le 1.4$ for 14 degrees of freedom,
corresponding to a probability of $\ge 0.14$) of the pulse profiles
during the observations in 1999 (Fig. \ref{fig: 1999 ASCA pulse profile}).
However, the template pulse profiles in 1994 and in 1998 differed
significantly from that in 1999 (Fig. \ref{fig: 1994 1998 ASCA pulse profile}).
The reduced $\chi^2$ values were 3.8 and 5.9,
corresponding to the probabilities of $1.8 \times 10^{-6}$
and $9.3 \times 10^{-12}$.
The pulse profiles in 1994 and in 1998 were consistent for $\chi^2_\nu = 1.4$.

We examined the possibility of whether
these pulse profile changes were caused by 
various mixing of the high-bit and medium-bit rate mode data.
In fact, the ratio of the high-bit to medium-bit rate mode exposures (H/M)
ranged from 0.2 to 2.9.
We compared four pulse profiles created from the 1999 ASCA data
with the 1999 ASCA template pulse profile to determine
how H/M variation affects the pulse profile.
These four pulse profiles were created from data in which
H/M = infinity (H = 0), H/M = 2.8, H/M = 1.2, and H/M = 0.
The reduced $\chi^2$ values of the pulse profile differences were
1.8, 1.0, 0.6, and 1.4.
This investigation verifies that the H/M variation
does not significantly affect the pulse profile change.
 
We also searched for the energy dependence of the pulse profile changes.
We divided the energy range of 0.5 to 10 keV into a high energy band
(3.0 to 10.0 keV) and a low energy band (0.5 to 3.0 keV).
The threshold of 3.0 keV was determined by the spectrum of 4U 0142+61.
The photon count from the power-law component
becomes greater above this energy than that from the blackbody component
(see Fig. 2 in \citet{White et al 1996}).
We found that the pulse profile change was only significant
in the low energy band.
However, this does not signify that the pulse profile change
only occurred in the low energy band,
because the photon counts in the high energy band were
about 1/10 times lower than that in the low energy band.

\section{Discussion}

We discovered a significant
frequency deviation from the simple spin-down law
in 4U 0142+61 for the first time.
It is not likely that this deviation was
caused by the timing noise,
because the timing solutions were good
during the {\it RXTE} observations (``First'' and ``Second'' spans)
and the {\it ASCA} 1999 observation.
The most likely interpretation of this frequency jump is a glitch.
The increase of the frequency ($\Delta \nu/\nu$)
can be constrained, 
assuming that the glitch happened between
    MJD 50893.00 (the last time of the First {\it RXTE} observation)
and MJD 51389.18 (the first time of the 1999 {\it ASCA} observation)
(Figure \ref{fig: Spin-down of 4U 0142+61}; lower panel).
The increase in the spin down rate can be estimated to be
$\Delta \dot{\nu}/\dot{\nu} = (1.4 \pm 0.5) \times 10^{-2}$
if the $\dot{\nu}$ difference in the {\it RXTE} ``First''
and ``Second'' ephemerides
(Table 2 in \cite{Gavriil-Kaspi 2002})
was caused by this glitch. 
Interestingly, these values are similar to those in a glitch of
another AXP 1RXS J170849.0$-$400910
(Glitch 1 in \cite{Kaspi-Gavriil 2003}).
These values are therefore also similar to those in the Vela radio pulsar
and other young radio pulsars 
(\cite{Kaspi-Lackey-Chakrabarty 2000} and references therein).

We also detected a significant pulse profile change 
in 4U 0142+61 for the first time,
which is possibly associated with the glitch.
This phenomenon is reminiscent of the out-burst that occurred in
another AXP 1E 2259+586 with regard to the following points
\citep{Kaspi 2003}.
(1) The amplitudes of the two peaks in the pulse profile were swapped
during the outburst in 1E 2259+586.
(2) 1E 2259+586 also underwent a sudden spin-up in the
magnitude of $\Delta \nu/\nu = 4 \times 10^{-6}$.
It could thus be interpreted that
4U 0142+61 occasionally underwent an outburst
as well as a pulse profile change and a glitch
just before the 1999 {\it ASCA} observation.
Nonetheless,
it must be noted that
the lack of a pulse profile change during the 1999 ASCA observation
suggests that survival time of the changed pulse profile is more than one month
and is longer than the considerably short-lived ($\sim 6$ days) pulse profile
change observed in 1E 2259+586.

The $\Delta \nu/\nu$ and $\Delta \dot{\nu}/\dot{\nu}$ 
of AXP glitches are similar to those of rotation-powered pulsars,
and therefore the inner structure of the neutron star
and the glitch mechanism may be basically the same.
Both types of neutron stars may consist of 
a neutron superfluid inner part
and an outer crust part,
and the glitch occurs due to the angular momentum transfer from
the inner superfluid to the crust
by a catastrophic vortex motion.
However, the pulse periods of AXPs ($\sim 10$ s) are 
beyond the cutoff of the glitch occurrence,
at which periods the stress of the Magnus force is not 
strong enough to trigger the glitch
(see Fig. 8 in \cite{Ruderman-Zhu-Chen 1998}).
This suggests that there is a difference in the mechanism
that causes these pulsars to reach the critical state of a glitch;
for example, the ordinary rotation-powered pulsars
reach the critical state due to their spin-down,
while the magnetar candidates do so
due to the stress of their magnetic fields.

The pulse profile change suggests two possibilities:
(1) Large scale reconnection of the magnetic field
or deformation of the crust
accompanied by deformation of the magnetosphere,
or
(2) Plasma ejection from the crust
into the magnetosphere (maybe a burst),
which changes the distribution of the plasma which emits
X-ray photons.
The association between the pulse profile change
and the glitch suggests the following scenarios:
(i) The glitch associated with the crust cracking
   \citep{Ruderman-Zhu-Chen 1998, Link-Epstein 1996}
   or platelet movement \citep{Ruderman 1991}
   causes deformation of the crust,
   and the pulse profile then changes as a result of (1) occurring
or
(ii) Plasma ejection from the crust associates
    with both the catastrophic vortex motion (a glitch)
    and the pulse profile changes due to the occurrence of (1) or (2).

We note that the observed ephemeris inconsistency 
can be explained without assuming a glitch.
It is possible that the spin-down rate
was temporarily smaller for a period between
the 1998 RXTE observation and the 1999 ASCA observations,
as observed in AXP 1E 1048.1$-$5937 \citep{Gavriil-Kaspi 2004}.
If the period of the smaller spin-down rate
begins at the end of the 1998 RXTE observation
and ends with the start of the 1999 ASCA observation,
the spin-down rate in that period becomes minimum.
It is $\dot{\nu} = -2.49(3) \times 10^{-14} $ Hz s$^{-1}$,
corresponding to the $5.9 \pm 1.2$\% change 
from the spin-down rate of the 1998 RXTE observation.
This lower limit of the $\dot{\nu}$ change is consistent
with the finding for AXP 1E 1048.1$-$5937 \citep{Gavriil-Kaspi 2004}.
Such behavior can be caused
by a change of the magnetospheric structure,
followed by a torque change and possibly followed by
the a pulse profile change.

\section{Conclusion}

An analysis of the {\it ASCA} observations
suggests the existence of a glitch in 4U 0142+61.
The glitch is similar to the glitch observed in
another AXP 1RXS J170849.0$-$400910 and other young radio pulsars.
This indicates that 
the radio pulsars and AXPs have the same
internal structures and glitch mechanisms.
However, the source of the stress
triggering the glitch may be different.
We also discovered a pulse profile change, possibly
associated with this glitch.
This suggests that AXPs have an association between
the internal structure and the magnetosphere
that does not exist in radio pulsars.

\acknowledgments

This work was partially supported by a 21st Century COE Program at
TokyoTech "Nanometer-Scale Quantum Physics" by the
Ministry of Education, Culture, Sports, Science and Technology.
This research was also supported in part by Grant-in-Aid for
Scientific Research(C)(15540239).

\end{document}